\documentclass[preprint,preprintnumbers, prd, floatfix, superscriptaddress,nofootinbib] {revtex4-1}
\usepackage{epsfig}
\usepackage{subfigure}
\usepackage{dcolumn}
\usepackage{bm}
\usepackage[usenames ,dvipsnames]{xcolor}
\usepackage{slashed}
\usepackage{graphicx,color}
\usepackage{hyperref}

\begin{document}

\title{Direct CP violation in\\internal $W$-emission dominated baryonic $B$ decays}

\author{Y.K. Hsiao}
\email{yukuohsiao@gmail.com}
\affiliation{School of Physics and Information Engineering, Shanxi Normal University, Linfen 041004, China}

\author{Shang-Yuu Tsai}
\email{shangyuu@gmail.com}
\affiliation{School of Physics and Information Engineering, Shanxi Normal University, Linfen 041004, China}

\author{Eduardo Rodrigues}
\email{eduardo.rodrigues@cern.ch}
\affiliation{Oliver Lodge Laboratory, University of Liverpool, Liverpool, UK}

\date{\today}
\begin{abstract}
The observation of CP violation has been experimentally verified
in numerous $B$ decays but is yet to be confirmed in final states with half-spin particles.
We focus our attention on baryonic $B$-meson decays mediated dominantly through internal $W$-emission processes
and show that they are promising processes to observe for the first time the CP violating effects
in $B$ decays to final states with half-spin particles.
Specifically, we study the $\bar B^0\to p\bar p\pi^0(\rho^0)$ and
$\bar B^0\to p\bar p\pi^+\pi^-$ decays.
We obtain ${\cal B}(\bar B^0\to p\bar p\pi^0)=(5.0\pm 2.1)\times 10^{-7}$,
in agreement with current data, and
${\cal B}(\bar B^0\to p\bar p\rho^0)\simeq {\cal B}(\bar B^0\to p\bar p\pi^0)/3$.
Furthermore, we find
${\cal A}_{CP}(\bar B^0\to p\bar p\pi^0,p\bar p\rho^0,p\bar p\pi^+\pi^-)
=(-16.8\pm 5.4,-12.6\pm 3.0,-11.4\pm 1.9)\%$.
With measured branching fractions ${\cal B}(\bar B^0\to p\bar p\pi^0,p\bar p\pi^+\pi^-)\sim {\cal O}(10^{-6})$,
we point out that ${\cal A}_{CP}\sim -(10-20)\%$
can be new observables for CP violation, accessible to the Belle~II and/or LHCb experiments.
\end{abstract}

\maketitle

\section{introduction}
The investigation of CP violation (CPV) has been one of the most important tasks
in hadron weak decays.
In the Standard Model (SM), CPV arises from a unique phase
in the Cabibbo-Kobayashi-Maskawa (CKM) quark mixing matrix;
however, it is insufficient to explain the matter and antimatter asymmetry of the Universe.
To try and shed light on solving the above puzzle,
a diverse set of observations related to CPV is necessary.
So far, direct CP violation has been observed in $B$
and $D$ decays~\cite{pdg,Aaij:2019kcg}.
With $Re(\epsilon'/\epsilon)$, it is also found in kaon decays~\cite{CPVinK}.
Although the decays involving half-spin particles offer an alternative route,
evidence for CP violation is not richly provided~\cite{Aaij:2014tua,Aaij:2016cla}.

Baryonic $B$ decays can be an important stage
to investigate CPV within the SM and beyond.
With $M^{(*)}$ denoting a pseudoscalar (vector) meson
such as $K^{(*)},\pi,\rho,D^{(*)}$, the $B\to p\bar p M^{(*)}$ decays
have been carefully studied
by the B factories and the LHCb experiment~\cite{Abe:2002ds, 
Aubert:2005gw,Aubert:2006qx,Chen:2008jy,Wei:2007fg,Aaij:2013fla,Aaij:2014tua}.
Experimental information includes measurements of branching fractions,
angular distribution asymmetries,
polarization of vector mesons in $B\to p\bar p K^*$,
Dalitz plot information,
and $p\bar p$ ($M^{(*)}p$) invariant mass spectra.
This helps to improve the theoretical understanding
of the di-baryon production
in $B\to{\bf B\bar B'}M$~\cite{Hou:2000bz,Geng:2006wz,
Hsiao:2016amt,Suzuki:2006nn,Hsiao:2018umx},
such that the data can be well interpreted.
Predictions are confirmed by recent measurements.
For example, one obtains
${\cal B}(\bar B^0_s\to p\bar \Lambda K^- + \Lambda\bar p K^+)
=(5.1\pm 1.1)\times 10^{-6}$~\cite{Geng:2016fdw},
in excellent agreement with the value of
$(5.46\pm 0.61\pm 0.57\pm 0.50\pm 0.32)\times 10^{-6}$ measured
by LHCb~\cite{Aaij:2017vnw}. Moreover,
the theoretical extension to four-body decays allows to interpret
${\cal B}(\bar B^0\to p\bar p \pi^+ \pi^-)$~\cite{Hsiao:2017nga,Aaij:2017pgn,Lu:2018qbw}.
The same can be said for CP asymmetries.

In this report we focus our attention on the baryonic $B$-meson decays
mediated dominantly through the internal $W$-emission diagrams.
Although the internal $W$-emission decays are regarded as suppressed processes,
the measured branching fractions of the baryonic $B$ decays
\begin{eqnarray}\label{belle_data}
{\cal B}(\bar B^0\to p\bar p\pi^0)
&=&(5.0\pm1.8\pm0.6)\times 10^{-7}\,,\nonumber\\
{\cal B}(\bar B^0\to p \bar p \pi^+\pi^-)
&=&(2.7\pm 0.1\pm 0.1\pm 0.2)\times 10^{-6}\,,
\end{eqnarray}
are not small~\cite{Pal:2019nvq,Aaij:2017pgn},
which make these modes an ideal place to observe for the first time
CP violation in $B$ decays to final states with half-spin particles.
Therefore, we will study the branching fractions
for the decays of $\bar B^0\to p\bar p \pi^0(\rho^0),\,p\bar p \pi^+\pi^-$,
and predict their direct CP violating asymmetries.

\section{Formalism}
\begin{figure}[t!]
\includegraphics[width=2.1in]{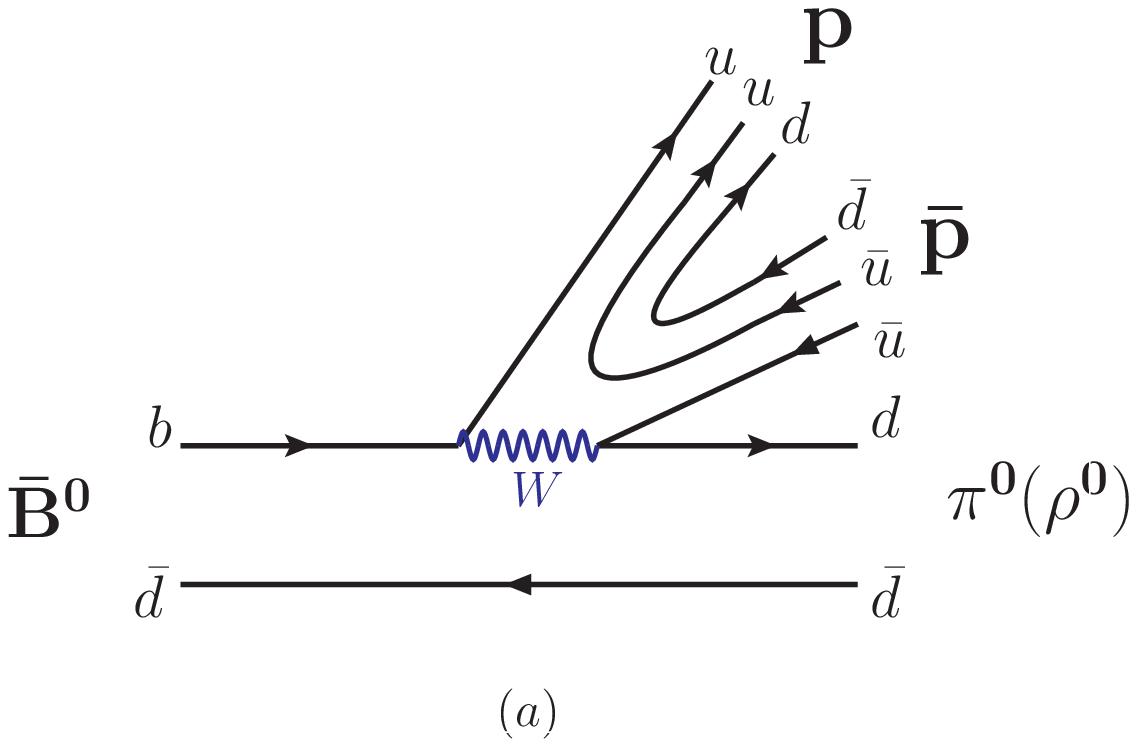}
\includegraphics[width=2.1in]{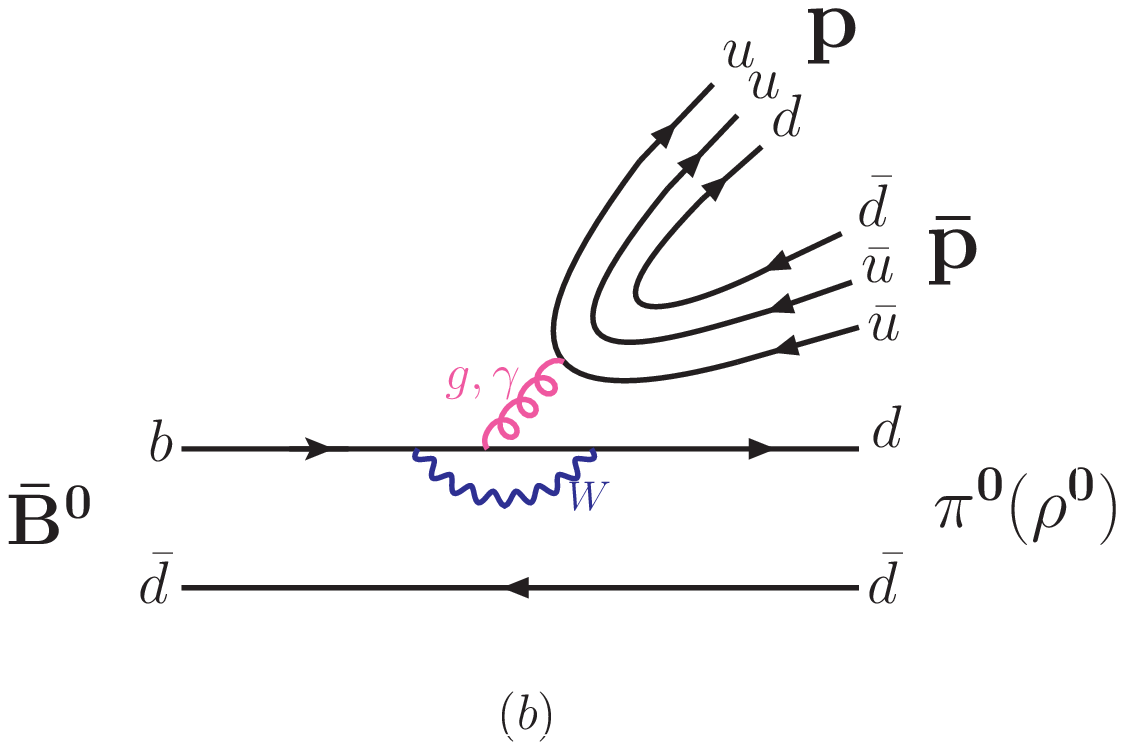}
\includegraphics[width=2.1in]{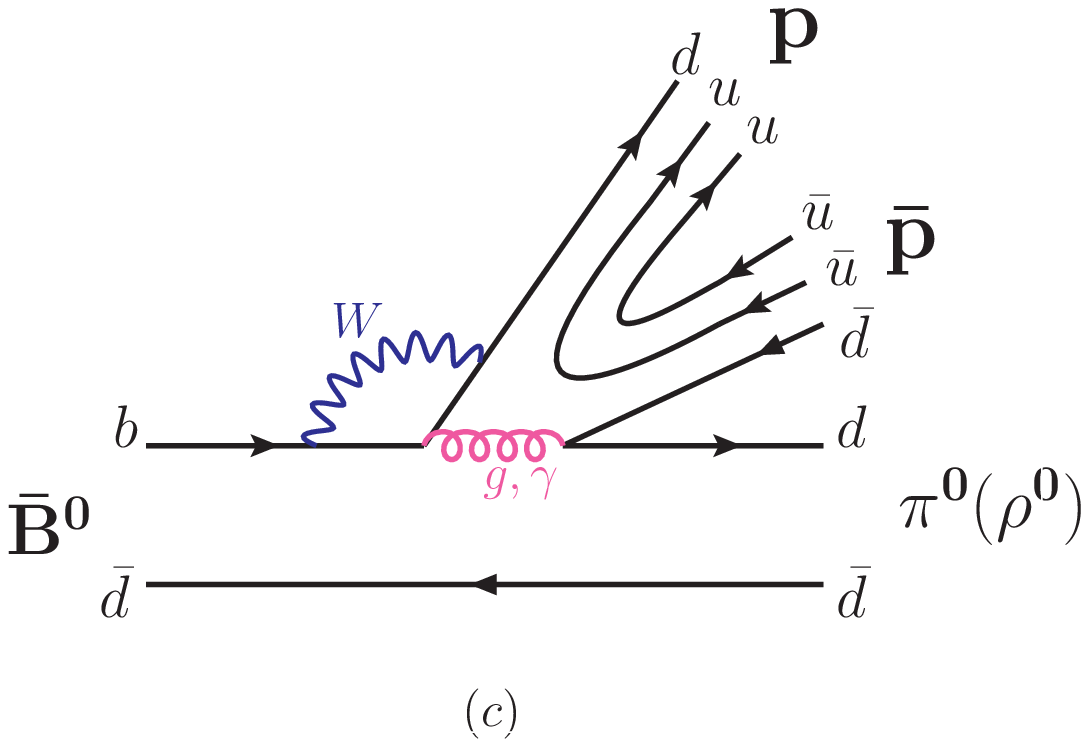}
\includegraphics[width=2.1in]{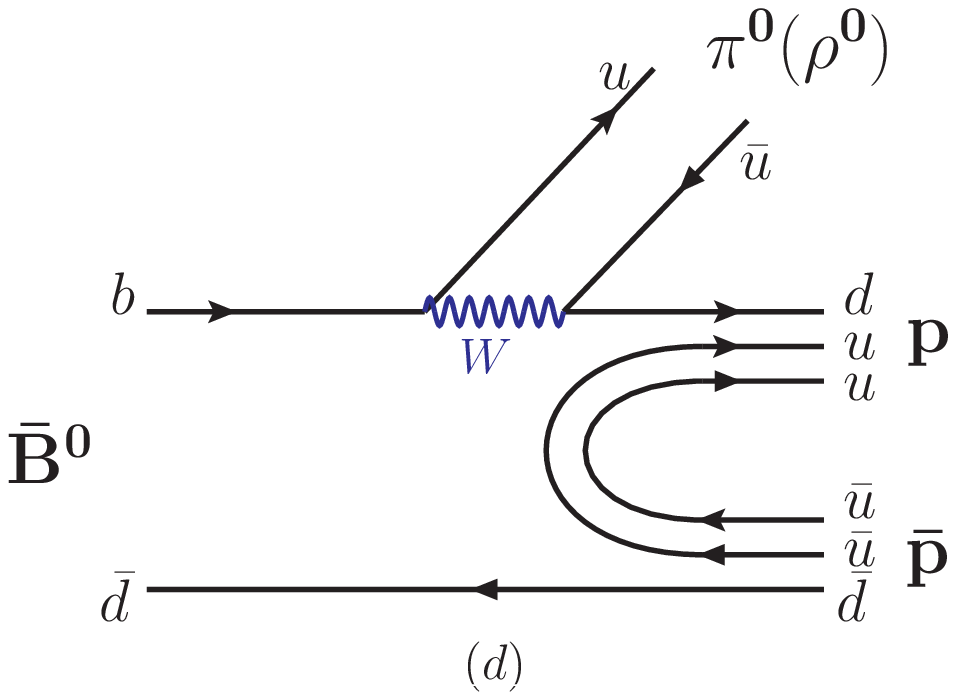}
\includegraphics[width=2.1in]{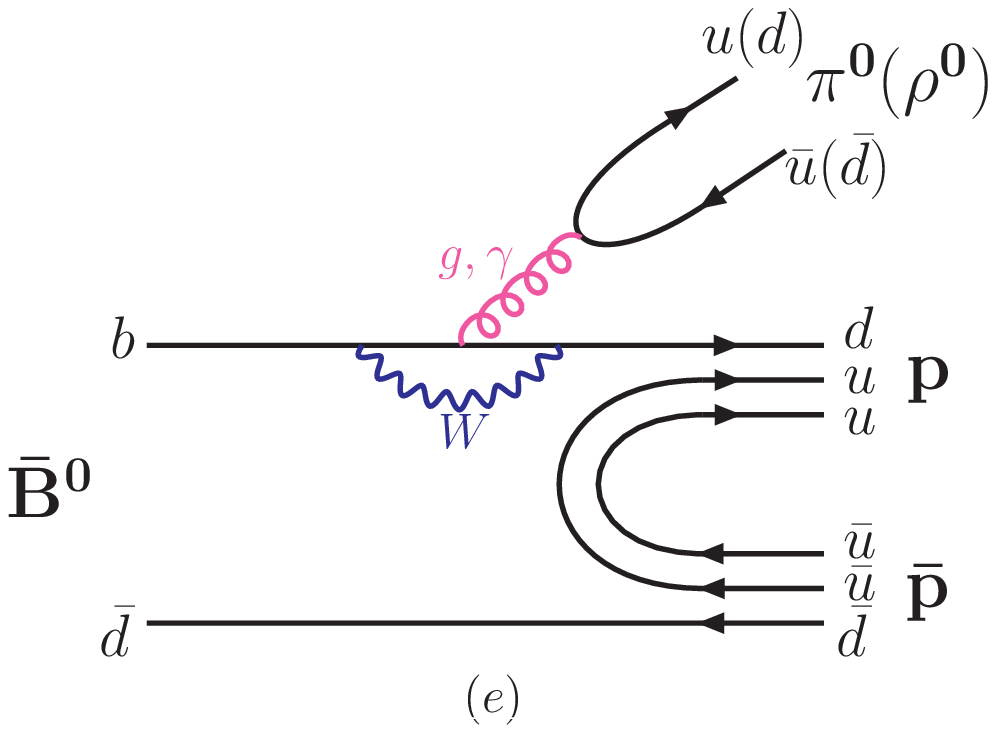}
\includegraphics[width=2.1in]{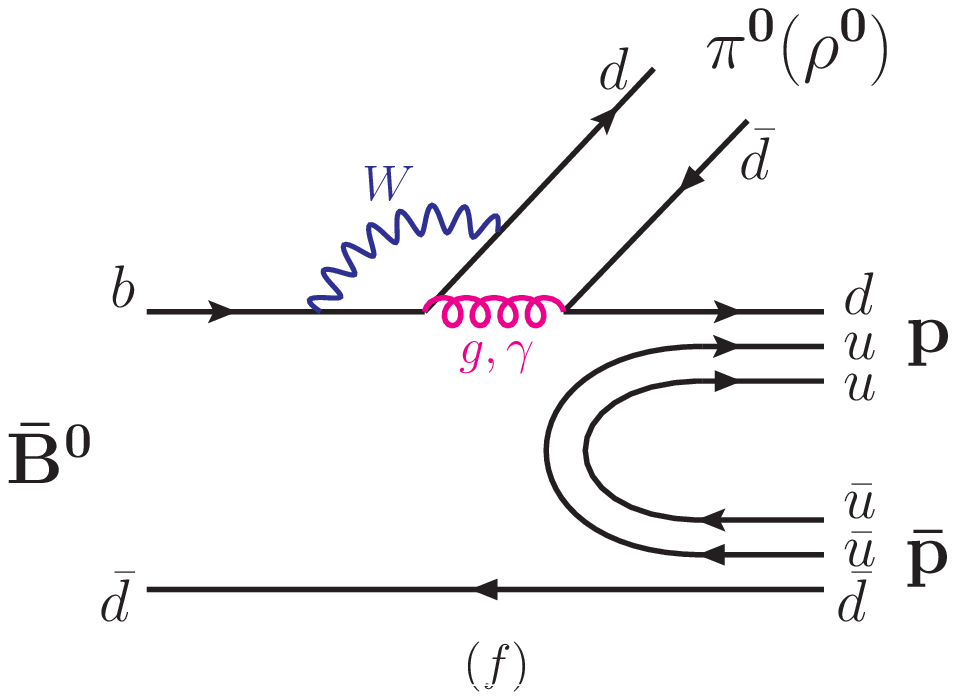}
\caption{The $\bar B^0\to p\bar p\pi^0(\rho^0)$ decay processes,
depicted as (a,b,c) for
the $\bar B^0\to\pi^0(\rho^0)$ transition with $0\to p\bar p$ production, and
(d,e,f) for
the $\bar B^0\to p\bar p$ transition with  recoiled $\pi^0(\rho^0)$ meson.}\label{dia}
\end{figure}
%
For the tree-level dominated $B$ meson decays,
the relevant effective Hamiltonian
is given by~\cite{Buras:1998raa}
\begin{eqnarray}\label{Heff}
{\cal H}_{eff}&=&\frac{G_F}{\sqrt 2}\bigg[ V_{ub}V_{ud}^*\bigg(\sum_{i=1,2} c_i O_i\bigg)
-V_{tb}V_{td}^*\bigg(\sum_{j=3}^{10} c_j O_j\bigg)\bigg]+h.c.,
\end{eqnarray}
where $G_F$ is the Fermi constant, 
$c_{i(j)}$ the Wilson coefficients, and
$V_{ij}$ the CKM matrix elements.
The four-quark operators $O_{i(j)}$ for the tree (penguin)-level contributions
are written as
\begin{eqnarray}\label{Oa}
&&
O_1=(\bar d_\alpha u_\alpha)_{V-A}(\bar u_\beta b_\beta)_{V-A}\,,\;
O_2=(\bar d_\alpha u_\beta)_{V-A}(\bar u_\beta b_\alpha)_{V-A}\,,\nonumber\\
&&
O_{3(5)}=(\bar d_\alpha b_\alpha)_{V-A}\sum_q(\bar q_\beta q_\beta)_{V\mp A}\,,\; 
O_{4(6)}=(\bar d_\alpha b_\beta)_{V-A}\sum_q(\bar q_\beta q_\alpha)_{V\mp A}\,,\nonumber\\
&&
O_{7(9)}={3\over 2}(\bar d_\alpha b_\alpha)_{V-A}\sum_q e_q(\bar q_\beta q_\beta)_{V\pm A}\,,\; 
O_{8(10)}={3\over 2}(\bar d_\alpha b_\beta)_{V-A}\sum_q e_q(\bar q_\beta q_\alpha)_{V\pm A}\,,
\end{eqnarray}
where $q=(u,d,s)$, 
$(\bar q_1 q_2)_{V\pm A}=\bar q_1\gamma_\mu(1\pm\gamma_5)q_2$, and 
the subscripts $(\alpha,\beta)$ denote the color indices.
With the identity of 
$\delta_{\beta\beta'}\delta_{\alpha\alpha'}
=\delta_{\alpha\beta}\delta_{\alpha'\beta'}/N_c+2T^a_{\alpha\beta}T^a_{\alpha'\beta'}$,
where $N_c=3$ is the color number, $O_i$ and $O_{i+1}$ can be related. 
For example, we have
$O_1=O_2/N_c+2\bar d\gamma_\mu(1-\gamma_5)T^a u\bar u\gamma^\mu(1-\gamma_5)T^a b$
with $T^a$ the Gell-Mann matrices.

\begin{figure}[t!]
\includegraphics[width=2.1in]{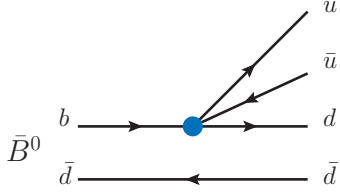}
\caption{The tree-level $b\to u\bar ud$ weak transition, where 
the blue blob represents the short-distance internal $W$-boson emission.}\label{uudd}
\end{figure}
%
In the factorization ansatz~\cite{Bauer:1986bm,ali}, one is able to express 
$\langle h_1 h_2|O|B\rangle$ as a product of two factors,
$\langle h_1 |J_1|0\rangle$ and $\langle h_2 |J_2|B\rangle$, 
where $O=J_1\cdot J_2$ is the product of the two color singlet quark currents $J_1$ and $J_2$
and $h_{1,2}$ denote the hadron states. The matrix elements
$\langle h_1 |J_1|0\rangle$ and $\langle h_2 |J_2|B\rangle$ are obtained in such a way
that the flavor quantum numbers of $J_{1,2}$ match the hadron states in the
separate matrix elements.
We hence decompose
$\langle p\bar p \pi^0|O_2|\bar B^0\rangle$ as~\cite{Hsiao:2018umx,Hsiao:2016amt}
\begin{eqnarray}
\langle O_2\rangle_{a}=\langle \pi^0|(\bar u_\beta u_\beta)_{V-A}|0\rangle
\langle p\bar p|(\bar d_\alpha b_\alpha)_{V-A}|\bar B^0\rangle\,,\nonumber\\
\langle O_2\rangle_{d}=\langle p\bar p|(\bar u_\beta u_\beta)_{V-A}|0\rangle
\langle \pi^0|(\bar d_\alpha b_\alpha)_{V-A}|\bar B^0\rangle\,, 
\end{eqnarray}
where the Fierz reordering has been used to exchange $(\bar d_\alpha,\bar u_\beta)$. 
The amplitudes
$\langle O_2\rangle_{a,d}$ correspond to the two configurations
depicted in Fig.~\ref{dia}(a,d), respectively. 
As depicted in Fig.~\ref{uudd} for the $b\to u\bar u d$ transition,  
dynamically, the $d$-quark moves collinearly
with the spectator quark $\bar d$ from $\bar B^0(b\bar d)$, so that
in Fig.~\ref{dia}(d) the $d\bar d$ for the $p\bar p$ formation can be seen as 
a consequence of the $B$ meson transition,
which is in accordance with the matrix element of 
$\langle p\bar p|(\bar d b)|\bar B^0\rangle$.
Moreover, since
$u\bar u$ and $d\bar d$ in the $\bar B^0$ rest frame can be seen to
move in opposite directions, we take $\pi^0(u\bar u)$ 
in Fig.~\ref{dia}(d) as the recoiled state,
in accordance with $\langle \pi^0|(\bar u u)|0\rangle$ 
with $|0\rangle$ representing the vacuum.
On the other hand, $\langle p\bar p \pi^0|O_1|\bar B^0\rangle$ is expressed as 
$\langle O_1\rangle_{a(d)}=\langle O_2\rangle_{a(d)}/N_c+\langle \chi_1\rangle$
with $\langle \chi_1\rangle\equiv 
\langle p\bar p \pi^0|2\bar u\gamma_\mu(1-\gamma_5)T^a d
\bar u\gamma^\mu(1-\gamma_5)T^a b|\bar B^0\rangle$. 
The $T^a$ in $\langle \chi_1\rangle$ correspond to
the gluon exchange between the two currents,
which causes an inseparable connection between the final states. Hence,
$\langle \chi_1\rangle$ is regarded as the non-factorizable QCD corrections.
Subsequently, we note that
$\langle p\bar p \pi^0|c_1 O_1+c_2 O_2 |\bar B^0\rangle=
a_2 \langle O_2\rangle_{a,d}$ with $a_2=c_2^{eff}+c_1^{eff}/N_c$,
where $c_i^{eff}$ represents the effective Wilson coefficient for $c_i$ 
to receive the next-to-leading-order contributions~\cite{ali}.
In the generalized edition of the factorization, one varies $N_c$ 
between 2 and infinity
in order to estimate $\langle \chi_1\rangle$~\cite{Bauer:1986bm,ali,Hsiao:2016amt}.
This makes $N_c$ a phenomenological parameter determined by data.

To complete the amplitudes,
we extend our calculation for $\langle p\bar p \pi^0|c_1 O_1+c_2 O_2 |\bar B^0\rangle$
to the penguin-level diagrams,
as depicted in Fig.~\ref{dia}(b,c,e,f). Moreover,
with $\pi^0$ replaced by $\rho^0$ and $\pi^+\pi^-$, we get 
the amplitudes of $\bar B^0\to p\bar p\rho^0$ and $\bar B^0\to p\bar p\pi^+\pi^-$,
respectively.
Hence, the decay amplitudes of $\bar B^0\to p\bar p X_M$
with $X_M\equiv (\pi^0(\rho^0),\pi^+\pi^-)$
can be written as~\cite{Hsiao:2018umx,Geng:2016fdw,Hsiao:2017nga}
\begin{eqnarray}\label{amp1a}
{\cal A}(\bar B^0\to p\bar p X_M)&=&{\cal A}_1(X_M)+{\cal A}_2(X_M)\,,
\end{eqnarray}
with ${\cal A}_{1,2}(X_M)$
corresponding to Fig.~\ref{dia}(a,b,c) and (d,e,f),
respectively. Explicitly, ${\cal A}_{1,2}$ are
given by~\cite{ali,Chua:2002wn,Chua:2002yd,
Hsiao:2016amt,Geng:2016fdw,Hsiao:2018umx}
\begin{eqnarray}\label{amp1b}
{\cal A}_1(X_M)&=&
\frac{G_F}{\sqrt 2}\bigg\{
\bigg[
\langle  p\bar p|\bar u\gamma^\mu(\alpha_2^+ -\alpha_2^-\gamma_5) u|0\rangle+
\langle  p\bar p|\bar d\gamma^\mu(\alpha_3^+ -\alpha_3^-\gamma_5) d|0\rangle\bigg]\nonumber\\
&\times& \langle X_M|\bar d \gamma_\mu(1-\gamma_5) b|\bar B^0\rangle
+\alpha_6\langle p\bar p|\bar d(1+\gamma_5) d|0\rangle
\langle X_M|\bar d(1-\gamma_5) b|\bar B^0\rangle\bigg\}\;,\nonumber\\
%
{\cal A}_2(X_M)&=&\frac{G_F}{\sqrt 2}\bigg\{
\bigg[\langle X_M|\bar u\gamma^\mu(\alpha_2^+ -\alpha_2^-\gamma_5) u|0\rangle+
\langle X_M|\bar d\gamma^\mu(\alpha_3^+ -\alpha_3^-\gamma_5) d|0\rangle\bigg]\nonumber\\
&\times& \langle p\bar p|\bar d \gamma_\mu(1-\gamma_5) b|\bar B^0\rangle 
+
\alpha_6\langle X_M|\bar d(1+\gamma_5) d|0\rangle
\langle p\bar p|\bar d(1-\gamma_5) b|\bar B^0\rangle\bigg\}\,.
\end{eqnarray}
%
The parameters $\alpha_i$
are defined as
\begin{eqnarray}\label{alpha_beta}
\alpha_2^\pm&=&
V_{ub}V_{ud}^* a_2-V_{tb}V_{td}^* (a_3\pm a_5\pm a_7+a_9)\,,\nonumber\\
\alpha_3^\pm&=&
-V_{tb}V_{td}^*(a_3+a_4\pm a_5\mp\frac{a_7}{2}-\frac{a_9}{2}-\frac{a_{10}}{2})\,,\nonumber\\
\alpha_6&=&V_{tb}V_{td}^*(2a_6-a_8)\,,
\end{eqnarray}
with $a_i\equiv c^{eff}_i+c^{eff}_{i\pm 1}/N_c$ for $i=$odd (even)~\cite{ali}.
We note that ${\cal A}_2(\pi^+\pi^-)$ is neglected since
${\cal A}_1(\pi^+\pi^-)\gg {\cal A}_2(\pi^+\pi^-)$~\cite{Hsiao:2017nga}.

The $B\to X_M$ transition matrix elements in
${\cal A}_1(X_M)$ are written as~\cite{BSW,Lee:1992ih}
\begin{eqnarray}\label{BtoM}
&&\langle M(p)|\bar q\gamma_\mu b|B(p_B)\rangle=\bigg[(p_B+p)^\mu-\frac{m^2_B-m^2_M}{q^2}q^\mu\bigg]
F_1^{BM}+\frac{m^2_B-m^2_M}{q^2}q^\mu F_0^{BM}\,,\nonumber\\
&&\langle M^*(p)|\bar q \gamma_\mu b|B(p_B)\rangle=\epsilon_{\mu\nu\alpha\beta}
\varepsilon^{\ast\nu}p_{B}^{\alpha}p^{\beta}\frac{2V_1}{m_{B}+m_{M^*}}\;,\nonumber\\
&&\langle M^*(p)|q\gamma_\mu \gamma_5 b|B(p_B)\rangle
=i\bigg[\varepsilon^\ast_\mu-\frac{\varepsilon^\ast\cdot
q}{q^2}q_\mu\bigg](m_B+m_{M^*})A_1
 + i\frac{\varepsilon^\ast\cdot q}{q^2}q_\mu(2m_{M^*})A_0\nonumber\\
&&\hspace{12em}-i\bigg[(p_B+p)_\mu-\frac{m^2_{B}-m^2_{M^*}}{q^2}q_\mu \bigg]
(\varepsilon^\ast\cdot q)\frac{A_2}{m_{B}+m_{M^*}}\;,\nonumber\\
&&\langle M_1(p_1) M_2(p_2)|\bar q\gamma_\mu(1-\gamma_5)b|B(p_B)\rangle
=\epsilon_{\mu\nu\alpha\beta}p_B^\nu (p_2+p_1)^\alpha (p_2-p_1)^\beta h \nonumber\\
&&\hspace{16em}
+iw_+ (p_2+p_1)_\mu+iw_-(p_2-p_1)+irq_\mu\,,
\end{eqnarray}
%
where
$\varepsilon_\mu$ is the polarization vector of $M^*$,
$q_\mu=(p_B-p)_\mu=(p_B-p_1-p_2)_\mu$
as the momentum transfer for the $B\to X_M$ transition,
$(F^{BM}_{0,1},V_1,A_{0,1,2})$ the $B\to M^{(*)}$ transition form factors
and $(h,r,w_{\pm})$ the $B\to M_1 M_2$ transition form factors.

The matrix elements of $0\to {\bf B\bar B'}$
are expressed as~\cite{Chua:2002yd}
\begin{eqnarray}\label{FFactor1}
\langle {\bf B\bar B'}|\bar q\gamma_\mu q'|0\rangle
&=&
\bar u\bigg[F_1\gamma_\mu+\frac{F_2}{m_{\bf B}+m_{\bf \bar B'}}i\sigma_{\mu\nu}q^\nu\bigg]v\;,\nonumber\\
\langle {\bf B\bar B'}|\bar q\gamma_\mu \gamma_5 q'|0\rangle
&=&\bar u\bigg[g_A\gamma_\mu+\frac{h_A}{m_{\bf B}+m_{\bf \bar B'}}q_\mu\bigg]\gamma_5 v\,,\nonumber\\
\langle {\bf B\bar B'}|\bar q q'|0\rangle &=&f_S\bar uv\;,
\langle {\bf B\bar B'}|q\gamma_5 q'|0\rangle =g_P\bar u \gamma_5 v\,,
\end{eqnarray}
where $u$($v$) is the (anti-)baryon spinor, and
$F_{1,2}$, $g_A$, $h_A$, $f_S$, $g_P$ the timelike baryonic form factors.

In ${\cal A}_2(\bar B^0\to p\bar p M^{(*)0})$,
the $0\to M^{(*)}$ matrix elements
are written as~\cite{pdg}
\begin{eqnarray}
\langle M(p)|\bar q\gamma_\mu \gamma_5 q'|0\rangle&=&-if_M p_\mu\,,
\langle M^*|\bar q\gamma_\mu q'|0\rangle=m_{M^*} f_{M^*}\varepsilon_\mu^*\,,
\end{eqnarray}
with $f_{M^{(*)}}$ the decay constant.
For the $B\to{\bf B\bar B'}$ transitions
we have~\cite{Chua:2002wn,Geng:2006wz}
\begin{eqnarray}\label{FFactor2}
\langle {\bf B\bar B'}|\bar q\gamma_\mu b|B\rangle&=&
i\bar u[  g_1\gamma_{\mu}+g_2i\sigma_{\mu\nu}\hat p^\nu +g_3 \hat p_{\mu}
+g_4(p_{\bf\bar B'}+p_{\bf B})_\mu +g_5(p_{\bf\bar B'}-p_{\bf B})_\mu]\gamma_5v\,,\nonumber\\
\langle {\bf B\bar B'}|\bar q\gamma_\mu\gamma_5 b|B\rangle&=&
i\bar u[ f_1\gamma_{\mu}+f_2i\sigma_{\mu\nu}\hat p^\nu +f_3 \hat p_{\mu}
+f_4(p_{\bf\bar B'}+p_{\bf B})_\mu +f_5(p_{\bf\bar B'}-p_{\bf B})_\mu]v\,,\nonumber\\
\langle {\bf B\bar B'}|\bar q b|B\rangle&=&
i\bar u[ \bar g_1\slashed {\hat p}+\bar g_2(E_{\bf \bar B'}+E_{\bf B})
+\bar g_3(E_{\bf \bar B'}-E_{\bf B})]\gamma_5v\,,\nonumber\\
\langle {\bf B}{\bf\bar B'}|\bar q\gamma_5 b|B\rangle&=&
i\bar u[ \bar f_1\slashed {\hat p}+\bar f_2(E_{\bf \bar B'}+E_{\bf B})
+\bar f_3(E_{\bf \bar B'}-E_{\bf B})]v\,,
\end{eqnarray}
where $\hat p_\mu=(p_B-p_{\bf B}-p_{\bf\bar B'})_\mu$,
$g_i(f_i)$ $(i=1,2, ...,5)$ and $\bar g_j(\bar f_j)$ $(j=1,2,3)$
are the $B\to{\bf B\bar B'}$ transition form factors.

The mesonic and baryonic form factors
have momentum dependencies.
For $B\to M^{(*)}$, they are given by~\cite{MFD}
\begin{eqnarray}\label{form2}
F_A(q^2)=
\frac{F_A(0)}{(1-\frac{q^2}{M_A^2})(1-\frac{\sigma_{1} q^2}{M_A^2}+\frac{\sigma_{2} q^4}{M_A^4})}\,,\;
F_B(q^2)=
\frac{F_B(0)}{1-\frac{\sigma_{1} q^2}{M_B^2}+\frac{\sigma_{2} q^4}{M_B^4}}\,,
\end{eqnarray}
where $F_A=(F_1^{BM},V_1,A_0)$ and $F_B=(F_0^{BM},A_{1,2})$.
According to the approach of perturbative QCD counting rules,
one presents the momentum dependencies of
the form factors for \mbox{$B\to{\bf B\bar B'}$,} $0\to{\bf B\bar B'}$ and
$B\to M_1 M_2$ as~\cite{Brodsky:1973kr,Brodsky:2003gs,
Chua:2002wn,Geng:2006wz,Chua:2002pi,Chua:2004mi}
\begin{eqnarray}\label{timelikeF2}
&&F_1=\frac{\bar C_{F_1}}{t^2}\,,\;g_A=\frac{\bar C_{g_A}}{t^2}\,,\;
f_S=\frac{\bar C_{f_S}}{t^2}\,,\;g_P=\frac{\bar C_{g_P}}{t^2}\,,\;\nonumber\\
&&f_i=\frac{D_{f_i}}{t^3}\,,\;g_i=\frac{D_{g_i}}{t^3}\,,\;
\bar f_i=\frac{D_{\bar f_i}}{t^3}\,,\;\bar g_i=\frac{D_{\bar g_i}}{t^3}\,,\nonumber\\
&&h=\frac{C_h}{s^2}\,,\;w_-=\frac{D_{w_-}}{s^2}\,,
\end{eqnarray}
where $t\equiv (p_{\bf B}+p_{\bf\bar B'})^2$, $s\equiv (p_1+p_2)^2$,
and $\bar C_i=C_i [\text{ln}({t}/{\Lambda_0^2})]^{-\gamma}$
with $\gamma=2.148$ and $\Lambda_0=0.3$ GeV.
In Ref.~\cite{Belitsky:2002kj}, $F_2=F_1/(t\text{ln}[t/\Lambda_0^2])$
is calculated to be much less than $F_1$; hence we neglect it.
Since $h_A$ corresponds to the smallness of
${\cal B}(\bar B^0\to p\bar p)\sim 10^{-8}$~\cite{Aaij:2013fta,Aaij:2017gum,Hsiao:2014zza},
we neglect $h_A$ as well.
The terms $(r,w_+)$ in Eq.~(\ref{BtoM}) are neglected
-- following~Refs.~\cite{Chua:2002pi,Chua:2004mi} --
due to the fact that their parity quantum numbers disagree with
the experimental evidence of $J^P=1^-$ for the meson-pair production~\cite{Drutskoy:2002ib}.

The constants $C_i$ ($D_i$) can be decomposed into sets of parameters
that obey the $SU(3)$ flavor and $SU(2)$ spin symmetries.
In Refs.~\cite{Brodsky:1973kr,Chua:2002yd,Hsiao:2017nga}
and \cite{Chua:2002wn,Geng:2006wz,Geng:2006jt,
Hsiao:2016amt,Geng:2016fdw}, they are derived as
\begin{eqnarray}\label{C&D}
&&
(C_{F_1},C_{g_A})=\frac{1}{3}
(5C_{||}+C_{\overline{||}},5C_{||}^*-C_{\overline{||}}^*)\,,\;
\text{(for $\langle p\bar p|(\bar u u)_{V,A}|0\rangle$)}\,\nonumber\\
&&
(C_{F_1},C_{g_A},C_{f_S},C_{g_P})=\frac{1}{3}
(C_{||}+2C_{\overline{||}},C_{||}^*-2C_{\overline{||}}^*,
\bar C_{||},\bar C_{||}^*)\,,\;
\text{(for $\langle p\bar p|(\bar d d)_{V,A,S,P}|0\rangle$)}\,\nonumber\\
&&
(D_{g_1,f_1},D_{g_j},D_{f_j})=\frac{1}{3}
(D_{||}\mp 2D_{\overline{||}},-D_{||}^j,D_{||}^j)\,,\;
\text{(for $\langle p\bar p|(\bar d d)_{V,A}|\bar B^0\rangle$)}\,\nonumber\\
&&
(D_{\bar g_1,\bar f_1},D_{\bar g_{2,3}},D_{\bar f_{2,3}})=\frac{1}{3}
(\bar D_{||}\mp 2\bar D_{\overline{||}},-\bar D_{||}^{2,3},-\bar D_{||}^{2,3})\,,\;
\text{(for $\langle p\bar p|(\bar d d)_{S,P}|\bar B^0\rangle$)}\,
\end{eqnarray}
with $j=2,..,4,5$,
$C_{||(\overline{||})}^*\equiv C_{||(\overline{||})}+\delta C_{||(\overline{||})}$ and
$\bar C_{||}^*\equiv \bar C_{||}+\delta \bar C_{||}$.
The direct CP violating asymmetry is defined as
\begin{eqnarray}\label{Acp}
{\cal A}_{CP}(B\to {\bf B\bar B'}X_M)\equiv
\frac
{\Gamma(B\to {\bf B\bar B'}X_M)-\Gamma(\bar B\to {\bf \bar B B'}\bar X_M)}
{\Gamma(B\to {\bf B\bar B'}X_M)+\Gamma(\bar B\to {\bf \bar B B'}\bar X_M)}\,,
\end{eqnarray}
where $\bar B\to {\bf \bar B B'}\bar X_M$ denotes
the anti-particle decay.

\begin{table}[tbhp]
\caption{The $\bar B^0\to M^{(*)0}$ transition form factors at zero-momentum transfer,
with $(M_A,M_B)=(5.32,5.32)$
and $(5.27,5.32)$~GeV for $\pi$ and $\rho$, respectively.}\label{MF}
\begin{tabular}{|c|cc|cccc|}
\hline
$\bar B^0\to \pi^0,\rho^0$&$F_1^{B\pi}$&$F_0^{B\pi}$ &$V_1$&$A_0$&$A_1$&$A_2$\\\hline
$\sqrt 2 f(0)$             &0.29 &0.29 &0.31 &0.30 &0.26 &0.24\\
$\sigma_1$ &0.48&0.76 &0.59 &0.54&0.73 &1.40\\
$\sigma_2$ &----- &0.28 &-----&-----&0.10 &0.50\\\hline
\end{tabular}
\end{table}

\section{Numerical Analysis}
We use the following values for the numerical analysis.
The CKM matrix elements are calculated via the Wolfenstein parameterization~\cite{pdg},
with the world-average values
\begin{eqnarray}\label{4_number}
\lambda=0.22453\pm 0.00044\,,
A=0.836\pm 0.015\,,
\bar \rho=0.122^{+0.018}_{-0.017}\,,
\bar \eta=0.355^{+0.012}_{-0.011}\,.
\end{eqnarray}
The decay constants are $f_{\pi,\rho}=(130.4\pm 0.2,210.6\pm 0.4)$~MeV~\cite{pdg},
with $(f_{\pi^0},f_{\rho^0})=(f_\pi,f_\rho)/\sqrt 2$.
We adopt the $B\to M^{(*)}$ transition form factors in Ref.~\cite{MFD}, listed in Table~\ref{MF}.
In Section~II, 
$N_c$ has been presented as the phenomenological parameter determined by data.
Empirically, one is able to determine $N_c$ between 2 and $\infty$.
With the nearly universal value for $N_c$ in the specific decays,
the factorization is demonstrated to be valid.
For the tree-level internal $W$-emission dominated $b$-hadron decays,
the extraction has given $N_c\simeq 2$ that corresponds to 
$a_2\sim {\cal O}(0.2-0.3)$~\cite{Hsiao:2016amt,Hsiao:2017nga,
Buras:1994ij,Neubert:2001sj,Hsiao:2017tif,Hsiao:2015txa,Hsiao:2015cda},
where $\delta N_c$ differs due to the experimental uncertainties.
For example, one obtains $N_c=2.15\pm 0.17$ 
in $\Lambda_b\to {\bf B}M_c$~\cite{Hsiao:2015txa,Hsiao:2015cda}.
Here, we test if $N_c\simeq 2$ 
can be used to explain the measured ${\cal B}(\bar B^0\to p\bar p\pi^0,p\bar p\pi^+\pi^-)$.

The $C_{h,w_-}$ for $\bar B^0\to \pi^+\pi^-$ 
and $C_i(D_i)$ for $0\to p\bar p$ ($\bar B^0\to p\bar p$) 
have been determined to be~\cite{Hsiao:2016amt,Geng:2016fdw,Hsiao:2017nga}
\begin{eqnarray}\label{fitC&D}
&&
(C_h,C_{w_-})=(3.6\pm 0.3,0.7\pm 0.2)\;{\rm GeV}^3\,, \nonumber\\
&&
(C_{||},C_{\overline{||}},\bar C_{||})=
(154.4\pm 12.1,18.1\pm 72.2,537.6\pm 28.7)\;{\rm GeV}^{4}\,,\nonumber\\
&&
(\delta C_{||},\delta C_{\overline{||}},\delta\bar C_{||})=
(19.3\pm 21.6,-477.4\pm 99.0,-342.3\pm 61.4)\;{\rm GeV}^{4}\,,\nonumber\\
&&
(D_{||},D_{\overline{||}})=(45.7\pm 33.8,-298.2\pm 34.0)\;{\rm GeV}^{5}\,,\nonumber\\
&&
(D_{||}^2,D_{||}^3,D_{||}^4,D_{||}^5)
=(33.1\pm 30.7,-203.6\pm 133.4,6.5\pm 18.1,-147.1\pm 29.3)\;{\rm GeV}^{4}\,,\nonumber\\
&&
(\bar D_{||},\bar D_{\overline{||}},\bar D_{||}^2,\bar D_{||}^3)
=(35.2\pm 4.8,-38.2\pm 7.5,-22.3\pm 10.2, 504.5\pm 32.4)\;{\rm GeV}^{4}\,.
\end{eqnarray}
For $\alpha_i$ in Eq.~(\ref{alpha_beta}),
the effective Wilson coefficients $c^{eff}_i$
are calculated at the $m_b$ scale in the NDR scheme, see Ref.~\cite{ali}.
They are related to the size of the decay,
where the strong phases, together with the weak phase in $V_{ub}$ and $V_{td}$,
play the key role in ${\cal A}_{CP}$.

Our results for
the branching fractions and CP violating asymmetries
of $\bar B^0\to p\bar p X_M$ decays are summarized in Table~\ref{pre},
where we have
averaged the particle and antiparticle contributions
for the total branching fractions.

\begin{table}[b]
\caption{Decay branching fractions and direct CP asymmetries of
$\bar B^0\to p\bar p X_M$, where
the first errors come from the estimations of the non-factorizable effects,
the second ones from the uncertainties of the CKM matrix elements, and
the third ones from those of the decay constants and form factors.}\label{pre}
\begin{tabular}{|l||c|c|}
\hline
& our result
&data\\\hline
$10^{7}{\cal B}(\bar B^0\to p\bar p\pi^0)$
&$5.0\pm 1.9\pm 0.3\pm 0.9$ 
&$5.0\pm1.9$ \cite{Pal:2019nvq}\\
$10^{7}{\cal B}(\bar B^0\to p\bar p\rho^0)$
&$1.8\pm 1.1\pm 0.1\pm 0.4$ 
&---\\
$10^{6}{\cal B}(\bar B^0\to p\bar p\pi^+\pi^-)$
&$2.7\pm 0.2\pm 0.2\pm 0.7$ 
&$2.7\pm 0.2$~\cite{Aaij:2017pgn}\\\hline
${\cal A}_{CP}(\bar B^0\to p\bar p\pi^0)$
&$(-16.8\pm 4.8\pm 1.6\pm 1.8)\%$ 
&---\\
${\cal A}_{CP}(\bar B^0\to p\bar p\rho^0)$
&$(-12.6\pm 2.2\pm 1.2\pm 1.7)\%$ 
&---\\
${\cal A}_{CP}(\bar B^0\to p\bar p\pi^+\pi^-)$
&$(-11.4\pm 0.2\pm 1.2\pm 1.4)\%$ 
&---\\\hline
\end{tabular}
\end{table}

\section{discussions and conclusions}
The improved theoretical approaches
such as QCD factorization (QCDF) and soft-collinear effective theory
have been applied to 
two-body mesonic $B$ decays~\cite{a2_add,Beneke:2001ev,Bauer:2001yt}.
Hence, the non-factorizable corrections of order $1/N_c^n$ with $n=1,2$
have been considered by calculating the vertex corrections  
from the hard gluon exchange and the hard spectator scattering.
Unfortunately, there exist no similar approaches well applied to
the $B\to M_1M_2M_3$, ${\bf B\bar B'}M$ and ${\bf B\bar B'}MM'$ decays,
due to the wave functions of $B\to{\bf B\bar B'}(MM')$ not as clear as those of $B\to M$.
By varying $N_c$ from 2 to $\infty$,
one can still estimate the non-factorizable QCD effects
with the corrections of order $1/N_c$.  This relies on
the generalized factorization, demonstrated to work well
in $B\to M_1M_2M_3$, $B\to{\bf B\bar B'}$,
$B\to {\bf B\bar B'}M({\bf B\bar B'}MM')$, $B\to D\pi$ and
$\Lambda_b\to {\bf B}M(\Lambda_c^+\pi^-)$~\cite{Chua:2002pi,Chua:2004mi,Hsiao:2014zza,
BtoMMM,Chua:2002wn,LbtoBM,BtoDpi,Wise}.
We determine $N_c=(2.15\pm 0.20,1.90\pm 0.03)$ 
to interpret ${\cal B}(\bar B^0\to p\bar p\pi^0,p\bar p \pi^+\pi^-)$
with $\delta N_c$ receiving the experimental uncertainties, 
which are indeed close to $N_c\simeq 2$ 
used in $B\to {\bf B\bar B'}M$ and $\Lambda_b\to {\bf B}M_{(c)}$~\cite{Hsiao:2016amt,
Hsiao:2015txa,Hsiao:2015cda,Hsiao:2017tif}.

In Table~\ref{pre},
${\cal B}(\bar B^0\to p\bar p\pi^0)=5.0\times 10^{-7}$
receives the contributions from 
${\cal A}_1,{\cal A}_2$ and their interference, denoted by ${\cal A}_{1\times 2}$,
which give ${\cal B}(\bar B^0\to p\bar p\pi^0)={\cal B}_1+{\cal B}_2+{\cal B}_{1\times 2}$
with $({\cal B}_1,{\cal B}_2,{\cal B}_{1\times 2})=(3.82,0.33,0.85)\times 10^{-7}$.
The ${\cal B}_{1\times 2}>0$ indicates constructive interference between ${\cal A}_{1,2}$.
By adopting $N_c$ from $\bar B^0\to p\bar p\pi^0$,
we predict ${\cal B}(\bar B^0\to p\bar p\rho^0)$. We find
${\cal B}(\bar B^0\to p\bar p\rho^0)\approx{\cal B}(\bar B^0\to p\bar p\pi^0)/3$ with
$({\cal B}_1,{\cal B}_2,{\cal B}_{1\times 2})=(2.00,0.04,-0.24)\times 10^{-7}$.
The minus sign of ${\cal B}_{1\times 2}$ indicates destructive interference.

With the theoretical approach reasonably well established
for the branching fractions, one can have reliable predictions for CP violation.
For example,
${\cal A}_{CP}(B^-\to p\bar p M^{(*)-})$ with $M^{(*)-}=(K^{*-},K^-,\pi^-)$
were predicted as $(22\pm 4,6\pm 1,-6\pm 1)\%$~\cite{Geng:2006jt},
agreeing with the experimental values of
$(21\pm 16,2.1\pm 2.0\pm 0.4,-4.1\pm 3.9\pm 0.5)\%$~\cite{pdg,Aaij:2014tua}.
Here,
our predictions for ${\cal A}_{CP}(\bar B^0\to p\bar p\pi^0(\rho^0),p\bar p\pi^+\pi^-)$
are around $-(10-20)\%$.
With $\delta{\cal A}_{CP}$ denoting the uncertainty for ${\cal A}_{CP}$,
we present $\delta{\cal A}_{CP}\simeq (0.2-0.3){\cal A}_{CP}$,
which receives the theoretical uncertainties from
the non-factorizable strong interaction, CKM matrix elements,
form factors and decay constants.

Expressing the decay amplitude as ${\cal A}=T e^{i\delta_W}+Pe^{i\delta_S}$,
the CP asymmetry can be derived as
\begin{eqnarray}\label{Rcp}
{\cal A}_{CP}=\frac{2R\sin\delta_W \sin\delta_S}{1+2R\cos\delta_W\cos\delta_S+R^2}\,,
\end{eqnarray}
where $\delta_W$ and $\delta_S$ are the weak and strong phases arising from the
tree~$(T)$ and penguin~$(P)$-level contributions, and
the ratio $R\equiv P/T$ suggests that a more suppressed $T$ amplitude
is able to cause a more sizeable ${\cal A}_{CP}$.
Although $\bar B^0\to p\bar p X_M$
involves complicated amplitudes, the relation in Eq.~(\ref{Rcp})
can be used as a simple description for
${\cal A}_{CP}(\bar B^0\to p\bar p X_M)$.
Being external and internal $W$-emission decays,
$B^-\to p\bar p\pi^-$ and $\bar B^0\to p\bar p\pi^0$ proceed with
$a_1\sim {\cal O}(1.0)$ and $a_2\sim {\cal O}(0.2-0.3)$
in the tree-level amplitudes~\cite{Geng:2006jt}, respectively.
Consequently,
the more suppressed $T$ amplitude with $a_2$ causes more interfering effect with
the penguin diagrams, which corresponds to
$|{\cal A}_{CP}(\bar B^0\to p\bar p\pi^0)|>|{\cal A}_{CP}(B^-\to p\bar p\pi^-)|$.
In fact,
we predict $|{\cal A}_{CP}(\bar B^0\to p\bar p\pi^0)|=(16.8\pm 5.4)\%$,
which is three times larger than $|{\cal A}_{CP}(B^-\to p\bar p\pi^-)|$~\cite{Geng:2006jt}.
For the same reason, $|{\cal A}_{CP}(\bar B^0\to p\bar p\rho^0,p\bar p\pi^+\pi^-)|$
can be as large as $(10-20)\%$.

Since ${\cal B}(\bar B^0\to p\bar p\pi^0,p\bar p\pi^+\pi^-)$
are measured as large as $10^{-6}$, and well
explained by the theory, with the predicted $|{\cal A}_{CP}|>10\%$,
they become promising decays for measuring CP violation.
By contrast, $\bar B^0\to p\bar p\rho^0$
as well as the internal $W$-emission dominated
$\Lambda_b$ decays of $\Lambda_b^0\to n\pi^0,n\rho^0$
have ${\cal B}\simeq (1-2)\times 10^{-7}$, which
make CP measurements a challenge even
in the case of large $|{\cal A}_{CP}|>10\%$~\cite{Hsiao:2017tif}.

In summary, we have investigated the branching fractions and
direct CP violating asymmetries of the
$\bar B^0\to p\bar p\pi^0(\rho^0)$ and $\bar B^0\to p\bar p \pi^+\pi^-$ decays.
We have shown that these baryonic $B$-meson decays
mediated dominantly through internal $W$-emission processes
are promising processes to observe for the first time the CP violating effects in
$B$ decays to final states with half-spin particles.

With a large predicted CP asymmetry ${\cal A}_{CP}=(-16.8\pm 5.4)\%$,
which is accessible to the Belle II experiment,
$\bar B^0\to p\bar p\pi^0$ is particularly suited
for a potential first observation of CP violation in baryonic $B$ decays
in the coming years.
Furthermore, the $\bar B^0\to p\bar p\pi^+\pi^-$ decay,
with its branching fraction of order $10^{-6}$
and the large predicted direct CP asymmetry ${\cal A}_{CP}\sim -(10-20)\%$,
is also in the realm of both Belle II and LHCb experiments.

\section*{ACKNOWLEDGMENTS}
This work was supported in part by
National Science Foundation of China (11675030) and
U. S. National Science Foundation award ACI-1450319.

\end{document}